\documentclass[useAMS,usenatbib]{mn2e}
\usepackage{txfonts,fleqn,graphicx}

\title{Halo evolution in the presence of a disc bar}

\author[P.~J.~McMillan \& W.~Dehnen]{
  Paul J.~McMillan
  and  Walter Dehnen\\
  Department of Physics \& Astronomy, 
  University of Leicester, Leicester, LE1 7RH
}

\begin{document}

\maketitle

\begin{abstract}
  Angular momentum transfer from a rotating stellar bar has been
  proposed by \cite{WeinbergKatz2002} as a mechanism to destroy
  dark-matter cusps in a few rotation periods. The $N$-body
  simulations performed by these authors in support of their claim
  employed spherical harmonics for the force computation and were, as
  shown by \cite{Sellwood2003}, very sensitive to inclusion of
  asymmetric terms (odd $l,m$). In order to disentangle possible
  numerical artifacts due to the usage of spherical harmonics from
  genuine stellar dynamical effects, we performed similar experiments
  using a tree code and find that significant cusp destruction requires
  substantially more angular momentum than is realistically available.
  However, we find that the simplified model (a $N$-body halo torqued
  by a rotating bar pinned to the origin) undergoes an instability in
  which the cusp moves away from the origin.  In presence of this
  off-centring, spherical density profiles centred on the origin
  display an apparent cusp-removal. We strongly suspect that it is
  this effect which \citeauthor{WeinbergKatz2002} observed.  When
  suppressing the artificial instability, cusp removal is very slow
  and requires much more angular momentum to be transferred to the
  halo than a realistic stellar bar possibly possesses.
\end{abstract}


\section{Introduction} \label{sec:intro}
Cold Dark Matter (CDM) cosmological simulations consistently predict
that the density profile of dark matter halos increases sharply
towards the centre, creating a `cusp', where the density increases
like a inverse power-law. \cite*{NavarroFrenkWhite1996} proposed
\begin{equation} \label{eq:nfw}
  \rho(r)=\frac{\rho_{c}}{(r/r_{\mathrm{s}})(1+r/r_{\mathrm{s}})^{2}},
\end{equation}
as a universal density profile (hereafter called `NFW') for CDM haloes
of all masses. Here, $r_{\mathrm{s}}$ is a scale radius and $\rho_{c}$
is a characteristic density. This density profile has a central slope
of $\mathrm{d}\ln\rho/\mathrm{d}\ln r=-1$. Later work has questioned
this inner slope and the universality over all halo masses, but the
existence of a cusp with $\mathrm{d}\ln\rho/\mathrm{d}\ln r$ in the
range $-1$ to $-1.5$ is a robust prediction of such simulations
\citep[e.g.][]{MooreEtal1998, PowerEtal2003}.

Observations of the rotation curves of disk galaxies are the best
probes of dark matter on galactic scales, and while some studies
cannot rule out consistency with a cuspy NFW-like profile
\citep{SwatersEtal2003,vandenBoschSwaters2001}, the majority of the 
observations are
claimed to be inconsistent with the density profile model
\citep{CoteCarignanFreeman2000, BinneyEvans2001,
  deBlokMcGaughRubin2001, deBlokEtal2001,
  BlaisOuelletteAmramCarignan2001, SalucciWalterBorriello2003,
  GentileEtal2004}. Investigations combining rotation curves with
gravitational lensing observations came to the same conclusion
\citep{TrottWebster2002}. While it has been suggested that the
discrepancy may be a result of direct comparison with the NFW fitting
formula, rather than with more realistic triaxial halos
\citep{HayashiEtal2004}, it is considered the greatest current
challenge of the CDM paradigm.

Numerous possibilities have been put forward in an effort to explain
the difference between theory and observation. Many of these involve
altering the assumed properties of dark matter, either the dark matter
particle itself \citep{SpergelSteinhardt2000,
  KaplinghatKnoxTurner2000}, or the initial power spectrum on small
scales \citep[warm dark matter; e.g.][]{ColinAvilaReeseValenzuela2000,
  BodeOstrikerTurok2001}. Since the central region of most galaxies
are dominated by baryonic matter, it seems reasonable that the
interaction between baryonic and dark matter will have a major effect
upon the dark matter density in the inner region of the halo.  If the
effect of this interaction was to reduce the dark matter density at
the centre of the halo, then this could potentially solve the problem.
Several investigation of such processes have been undertaken, some
pointing out effects that would \emph{increase} the gradient of the
density cusp still further \citep{BlumenthalEtal1986}, some discussing
processes that could potentially remove the cusp
\citep{BinneyGerhardSilk2001, ElZantShlosmanHoffman2001,
  WeinbergKatz2002}. In this paper we investigate the mechanism
proposed by \citeauthor{WeinbergKatz2002}, of angular momentum
transfer from a rotating disc bar to the halo as a means of removing a
cusp.
  
\citeauthor{WeinbergKatz2002} pointed to an earlier study by
\cite{HernquistWeinberg1992} to suggest that angular momentum would be
rapidly transfered from a rotating disc bar to the halo at an inner
Lindblad-like resonance. \citeauthor{WeinbergKatz2002} supported the
hypothesis with analytical calculations and simplified $N$-body
simulations. In the simulations, the gravitational potential of a
rotating bar, with centre pinned at the initial centre of the density
distribution, was imposed upon an $N$-body representation of a CDM
halo with an NFW profile.

The simulations of \citeauthor{WeinbergKatz2002} were re-examined by
\cite{Sellwood2003} who found the same reduction in central density,
but argued that the deliberately artificial perturbation applied to
the system lead to misleading results. \cite{Sellwood2003} ran fully
self-consistent simulations, which showed a slight \emph{steepening}
of the halo's inner profile from the action of a bar. This result was
also seen in the simulations of \cite{ValenzuelaKlypin2003}. However 
the self consistent simulations of \cite{HolleyBockelmannWeinbergKatz2003} 
showed a clear flattening in the cusp, as do the later ones of 
\cite{Weinberg2004}. A discussion of this apparently irreconcilable 
discrepancy was given by \cite{Athanassoula2004}.

The simulations of \cite{WeinbergKatz2002} and \cite{Sellwood2003},
and also the original study of \cite{HernquistWeinberg1992} all
employed spherical harmonics to compute the gravitational potential
and accelerations. The main motivation for this choice was the high
symmetry of the problem, which allows the restriction to a low number
of harmonics. This in turn reduces the computational effort and
enables a larger number $N$ of bodies and hence higher numerical
resolution than with more traditional methods, such as the Barnes \&
Hut (\citeyear{BarnesHut1986}) tree code. In fact,
\cite{WeinbergKatz2002} even claim that cusp removal cannot be
successfully simulated by a tree code (unless $N\gg10^6$), because the
noise in the forces would scatter bodies off the resonant orbits,
which, according to these authors, are driving the angular momentum
transfer. Because of this argument, it was not entirely clear whether
the aforementioned results of fully self-consistent simulations are
realistic or not.

There is, however, a price to pay for using spherical harmonics. Most
importantly, any method based on spherical polar coordinates is
vulnerable to off-centring: if the modelled density distribution is
not centred on the origin of the coordinate system, many terms are
required for the series to converge.  Conversely, when off-centring
occurs, the gravity obtained from only the first few terms is strongly
biased, which may induce artificial $m=1$ instabilities and/or
interfere with natural $m=1$ instabilities. The only proper way to
avoid this problem is to enforce centring, i.e.\ not allowing
coefficients with odd $l,m$ to carry any weight. Clearly, this makes
systems with any properties that are not reflection symmetric,
including instabilities, impossible to simulate in this way.

\cite{WeinbergKatz2002} included coefficients of odd $(l,m)$ and hence
their simulations were vulnerable to artificial effects due to
off-centring. \cite{Sellwood2003} could reproduce their results, but
if he suppressed the coefficients with odd $(l,m)$, the results were
completely altered in that more than hundred instead of a few bar
rotations were required to remove the density cusp. Sellwood also
pointed to a previous study by \cite{White1983} which reported that
the $l=1$ term can cause numerical artifacts. Thus, apparently, the
simplified bar-halo model of \citeauthor{WeinbergKatz2002} undergoes a
non-reflection symmetric evolution. However, owing to the problems
inherent to the use of spherical harmonics, it is unclear from the
studies of \citeauthor{WeinbergKatz2002} and \cite{Sellwood2003}
whether this evolution is an artifact of their algorithm (i.e.\ the
usage of spherical harmonics) or not.

In an effort to resolve this question we performed similar simulations
ourselves with a different type of $N$-body code, which does not rely
on an expansion in spherical harmonics and hence avoids these problems.
Technical details of the models used, and the $N$-body code are given in 
\S\ref{sec:model}, our results are presented in \S\ref{sec:results}. 
We discuss our results and draw conclusions in \S\ref{sec:conclude}

\section{Modelling} \label{sec:model}
For the majority of our simulations we model the halo as an 
isotropic spherical \cite{Hernquist1990} model, which has density
\begin{equation} \label{eq:rho}
  \rho (r) = \frac{M_{\mathrm{halo}}\,r_{\mathrm{s}}}{2\pi
    r(r_{\mathrm{s}} + r)^{3}}
\end{equation}
with $M_{\mathrm{halo}}$ and $r_{\mathrm{s}}$ the total mass and scale
radius, respectively.  This profile has the same `cuspy' behaviour as
the NFW profile (\ref{eq:nfw}) in the limit $r\to0$, but has the
virtue of having less mass in the outer regions of the halo. This
means that a simulation with the same particle number will have
greater resolution in the cusp.  We introduce the bar as an external
potential applied to the $N$-body simulation. Only the quadrupole
moment of the bar is included in the force calculation, since the
monopole term would have the effect of adding mass at the centre of
the halo, altering the equilibrium, and higher order terms are far
weaker, having little effect on the evolution, and are ignored for
simplicity. A convenient fitting formula, which behaves properly as a
quadrupole for both $r\to 0$ and $r\to\infty$ is provided by
\cite{HernquistWeinberg1992}:
\begin{equation} \label{eq:bar}
  \Phi_{\mathrm{bar}} = -\frac{GM_{\mathrm{bar}}}{a}
  \frac{\alpha r^{2}_{\ast}}
  {(\beta^\gamma_{\phantom{\ast}} + r^\gamma_\ast)^{5/\gamma}}
  \sin^2\!\theta\,\cos2(\varphi-\Omega_{\mathrm{p}}t).
\end{equation}
Here, $r_\ast=r/a$ with $a$ the semi-major axis of the bar,
$r,\,\theta,$ and $\varphi$ are the usual spherical polar coordinates,
and $M_{\mathrm{bar}}$ and $\Omega_{\mathrm{p}}$ the mass and pattern
speed of the bar, respectively. $\alpha$, $\beta$ and $\gamma$ are
dimensionless parameters determined by best fit to the chosen bar
density profile and shape. The bar is `turned on' adiabatically over
$\sim10$ rotation periods to minimise transient effects.

In the majority of cases we follow \cite{HernquistWeinberg1992} 
and \cite{Sellwood2003} and model the bar as an $n=2$
\cite{Ferrers1877} bar with axis ratio $1:0.5:0.1$. This is fitted by
the parameters $\alpha\simeq0.1404$, $\beta\simeq0.4372$ and
$\gamma=2$  (while an incorrect version of this fitting formula is
reproduced by \cite{Sellwood2003}, the correct model was used in his
simulations, Sellwood, private communication).

The simulations of \cite{WeinbergKatz2002} were of a NFW density profile 
halo (\ref{eq:nfw}), truncated at large radii (as the NFW model has 
infinite mass, if considered for $r\to\infty$). The bar density profile 
used was that of a homogeneous ellipsoid of axis ratio $1:0.5:0.05$. This 
is well fitted by the
same fitting formula used for the n=2 Ferrers bar (\ref{eq:bar}), with the 
parameters $\alpha\simeq0.1227$, $\beta\simeq0.6288$ and $\gamma=5.314$. In
Section~\ref{sec:results:NFW} we present simulations with the same parameters 
as \cite{WeinbergKatz2002} in an effort to directly compare our simulations 
with theirs. For this we used a truncated NFW halo which we defined as 
having the density profile

\begin{equation} \label{eq:tNFW}
\rho (r) = \frac{\rho_{\mathrm{c}}}{(r/r_{\mathrm{s}})(1 + r/r_{\mathrm{s}})^{2}}   \textrm{sech}(r/10r_{\mathrm{s}})
\end{equation}

\subsection{Technical details} \label{sec:model:tech}
We generate the initial positions and velocities from the density
(\ref{eq:rho}) and the isotropic distribution function.
In the case of the Hernquist halo this is known \citep{Hernquist1990}, while 
in the case of the truncated NFW halo it can be found numerically using 
Eddington's \citeyear{Eddington1916} formula. Quasi-random numbers were 
employed in order to suppress
Poisson noise. The $N$-body simulations were performed using the
publicly available $N$-body code \textsf{gyrfalcON}, which is based on
Dehnen's (\citeyear{Dehnen2000:falcON, Dehnen2002}) force solver
\textsf{falcON}, a tree code with mutual cell-cell interactions and
complexity $\mathcal{O}(N)$. The performance of the code is comparable
to that of the spherical-harmonic based so-called SCF method
\citep{HernquistOstriker1992, Weinberg1999} with terms up to $n,l=8$
included.

For the Hernquist halo we used units of time, mass, and length such 
that $G\equiv1$,  $r_{\mathrm{s}}\equiv11$ and $M_{\mathrm{halo}}\equiv1$. 
For the truncated NFW halo we used units such that $G\equiv1$,  
$r_{\mathrm{s}}\equiv1$, and $v_{\mathrm{circ,max}}\equiv1$
The equations of motion were integrated using the familiar leap-frog 
integrator with minimum time step $2^{-7}$ and a block-step scheme allowing 
steps up to four times larger. Individual particle time steps were adjusted 
in an (almost) time-symmetric fashion such that on average
\begin{equation}
  \tau_i = \min\left\{\frac{0.01}{|\bmath{a}_i|},\;
    \frac{0.01}{|\Phi_i|}\right\}
\end{equation}
with $\Phi_i$ and $\bmath{a}_i$ the gravitational potential and
acceleration of the $i$th body. A fiducial simulation with $N=3 \times 10^5$
and no imposed bar conserved energy to within 0.2\% over 1500 time
units corresponding to $\sim$120 bar rotations.

Throughout this paper lengths will be quoted in terms of halo scale
lengths
.

\section{Results} \label{sec:results}
\begin{figure}
  \centerline{\resizebox{\hsize}{!}{\includegraphics{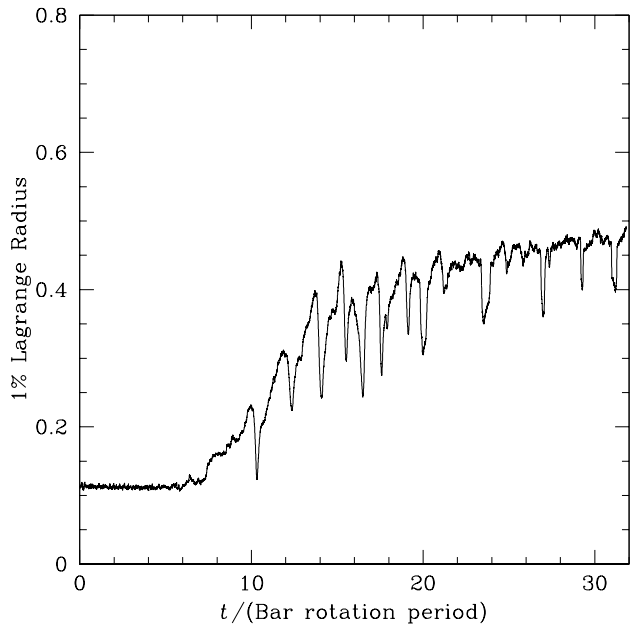}}}
  \caption{Time evolution of the 1\% Lagrange radius in flawed simulation.
    \label{fig:Rapidevo:Lagr}
  }
\end{figure}

\begin{figure}
  \centerline{\resizebox{\hsize}{!}{\includegraphics{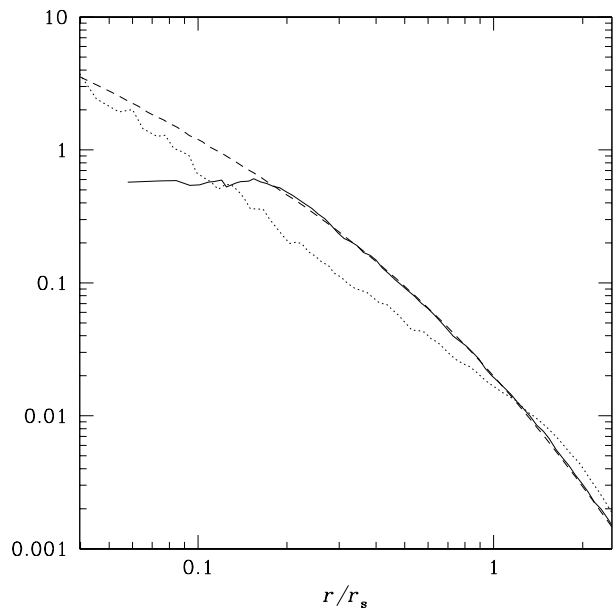}}}
  \caption{Spherically averaged density profile (inner part) for the same
    simulation as in Fig.~\ref{fig:Rapidevo:Lagr}. Shown are
    the initial profile (\emph{dashed}); and the profile after 8 bar 
    rotations, considered about the original centre (\emph{solid}), 
    or about the centre of the cusp (\emph{dotted}).  The
    profile appears to be cored after 8 bar rotation times but this is
    because the implicit assumption of conserved (or at least
    approximately conserved) spherical symmetry is incorrect, see
    Fig.~\ref{fig:Scatterplot}. The profile considered 
    about the centre of the cusp shows that the cusp is as strong 
    as ever, but the force applied by the bar has become wildly unphysical,
    so this is result is, in any case, insignificant.
    \label{fig:Rapidevo:rho}
  }
\end{figure}

Initial simulations of a Hernquist model halo used $N=300\,000$ 
and a bar of length (semi-major
axis) $a=0.7$ and mass 30\% of the halo mass interior to $r=a$, as it
was for all simulations in this halo type. The bar rotates with a fixed 
pattern speed
with co-rotation at $r_{\mathrm{s}}$, in this experiment and all
others.  These experiments showed a rapid change in the 1\% Lagrange
radius (the radius within which 1\% of the total mass of the halo is
contained, Fig.~\ref{fig:Rapidevo:Lagr}). This only demonstrates that the
density at the \emph{origin} was reduced by the action of the bar, which is not
necessarily the density peak of the halo.  The
spherically averaged density profile (Fig.~\ref{fig:Rapidevo:rho})
seems to show the formation of a core after just 8 bar rotations. Neither 
of these measures take into account any possible change in the position 
of the centre of the distribution.
Similar simulations with $N$ ranging from $10^4$ to $10^6$ showed
essentially the same behaviour.  This is in agreement with the apparent
rapid evolution seen in simulations including the $l=1$ term by 
\cite{Sellwood2003}, though the timescale is slightly longer. It is 
similar to the evolution observed by 
\cite{WeinbergKatz2002}, though the halo model is different 
(see Section ~\ref{sec:results:NFW}).


\subsection{Unstable evolution} \label{sec:results:unstable}
\begin{figure}
  \centerline{\resizebox{\hsize}{!}{\includegraphics{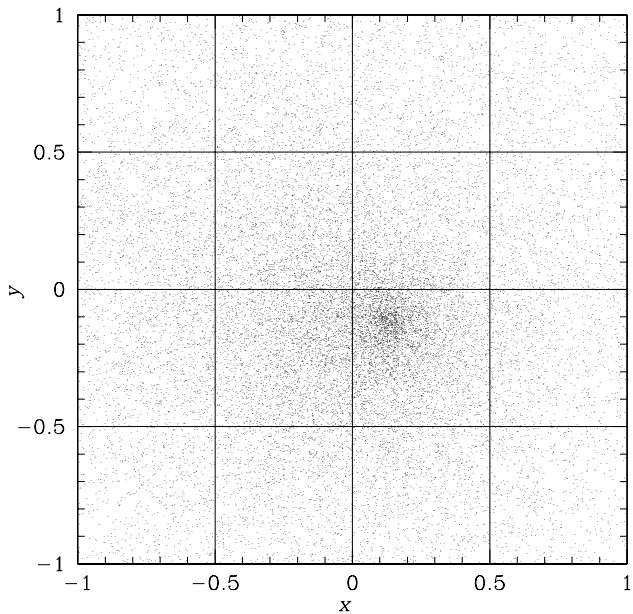}}}
  \caption{Positions in the $x$-$y$ plane of bodies with $|z|<1$ after 8
    bar rotations. The plot shows that the centre of the distribution
    has shifted from the origin, thus invalidating the simulation
    setup. \label{fig:Scatterplot} }
\end{figure}

This reduction in density is, however, not demonstrative of the
creation of a cored density profile. Simply plotting the particle
positions in the $x$-$y$ plane (Fig.~\ref{fig:Scatterplot}) shows that
within a few bar rotations the densest part of the halo is no longer
at the origin. This creates the appearance of a cored density profile
if analysis is performed with respect to the origin. This effect
occurs even if the initial distribution of bodies is made exactly
symmetrical initially (for each body at $\bmath{x},\bmath{v}$ there is
another one at $-\bmath{x},-\bmath{v}$).

This off-centring means that these simulation results are invalid. In this 
simple model, the
bar's centre is fixed to the origin and does not (is not allowed to)
react to the off-centring of the halo. This leads to a wildly
unphysical interaction between the bar and the halo which rapidly
transports energy and angular momentum to the halo, and invalidates
all results found this way. We strongly suspect that the rapid
evolution reported by \cite{WeinbergKatz2002} and its $l=1$ dependence
seen by \cite{Sellwood2003} are actually due to this off-centring. Since
this effect is reproduced in our simulations, it is clear that it is
a real instability of the simplified bar-halo system investigated,
and is not entirely due to an artifact of the spherical harmonic approach
noted by \cite{Sellwood2005}.

The initial conditions with the applied quadrupole potential are
entirely point symmetric about the origin. However they are unstable
to motion of the cusp away from the centre of the bar. This occurs
even when the initial particle distribution is defined to be exactly
symmetric, because slight asymmetries caused by numerical truncation
errors break the imposed symmetry.

\begin{figure}
  \centerline{\resizebox{\hsize}{!}{\includegraphics{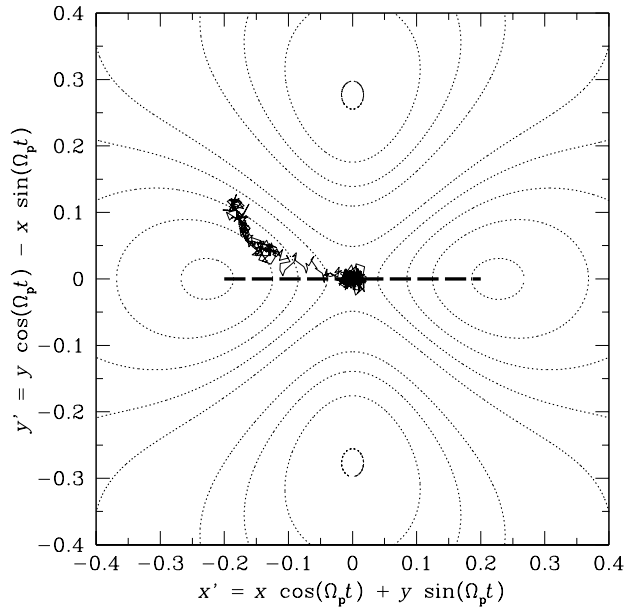}}}
  \caption{Motion of the density centre in the plane co-rotating with
    the bar during the first 11 bar rotations (\emph{solid line}) for
    the same simulation as in previous Figures. The \emph{thick dashed
      line} indicates the orientation of the bar, which rotates
    anticlockwise. Contours of the sum of the effective bar quadrupole 
    potential in this frame and an approximation for the halo-bulk 
    potential (halo without cusp, using a $\gamma=0$ Dehnen 
    (\citeyear{Dehnen1993}) model) are shown as \emph{dotted} curves. 
    Stationary points perpendicular to the bar are maxima, origin is 
    a saddle point, other stationary points along the bar axis are 
    minima. 
    \label{fig:quad}}
\end{figure}

The cause of this instability can be understood by considering the
external quadrupole potential applied to the $N$-body system.
Figure~\ref{fig:quad} shows the position of the central density in the
frame of the rotating bar, overlaid upon a contour plot of the
quadrupole potential. The cusp of the halo is initially at a saddle
point of the external potential. This is unstable to a small movement
of the cusp. We can think, in a simplistic picture, of the `cusp'
falling into a potential well. The restoring force due to the bulk of
the halo is too weak to counter this. We also note that the cusp does
not fall to the exact centre of the well, which is presumably due to 
the Coriolis force.

\subsection{NFW halo} \label{sec:results:NFW}
In an effort to re-examine the results of \cite{WeinbergKatz2002}, we ran 
simulations of an NFW halo with the same parameters they used. These 
simulations had a short, light bar with semi-major axis a=0.5$r_s$; and 
a mass equal to 15\% of the halo mass within the bar radius. Since 
\citeauthor{WeinbergKatz2002} suggested that high numerical resolution 
was required to accurately simulate the resonant dynamics, we modelled the 
halo with particles of varying mass, with the lowest mass particles in 
the inner regions where the important resonant processes occur. This 
allows an substantial increase in the resolution of the inner regions 
of the NFW halo, without increasing the total particle number. We can define 
$N_{eff}$, the effective particle number, as 
$N_{\textrm{eff}}=M_{\textrm{tot}}/m_{\textrm{lowest}}$, 
where $M_{\textrm{tot}}$ is the total mass of the halo, 
and $m_{\textrm{lowest}}$ is the mass 
of the lowest mass particles (the particles in the innermost regions of the 
halo). In this way we were able to improve the resolution of our simulations 
such that $N_{\textrm{eff}}\simeq 5\times N$ without loss of stability.

Experiments utilising gyrfalcON, and with the $N_{\textrm{eff}}$ 
varying between $30\,000$ and $6\,000\,000$ showed no significant evolution 
of the halo under these conditions. The bar potential was too weak to 
destabilise the halo and move its centre. This is in contrast to simulations 
with $N$ 
between $10\,000$ and $4\,000\,000$, run with the potential expansion codes 
of \citeauthor{WeinbergKatz2002}, and \citeauthor{Sellwood2003} which 
\emph{did} become unstable rapidly (within $\sim$10 bar rotations). The very 
low $N$ at which this instability is observable in the potential expansion 
case, and the absence of instability even in the high $N$ tree code 
(gyrfalcON) case, 
indicates that this difference is not caused by particle noise drowning out 
the effect of the bar in the tree code case. The potential expansion approach 
is clearly more susceptible to the instability of the system than a tree code.

It should be noted that there are two separate instabilities. The genuine 
instability of the artificial bar-halo system was responsible for the shift 
of the cusp away from the origin in the simulations with the Hernquist halo. 
In the NFW case the bar is too weak to destabilise the halo when gyrfalcON 
is used, but using a potential expansion approach exacerbates the problem 
because of the odd ($l,m$) instability referred to by \citeauthor{Sellwood2003}.  

\subsection{Preventing instability} \label{sec:results:stable}

\subsubsection{Forcing symmetry} \label{sec:results:stable:symm}
In our simplified model, both the halo and the bar are entirely point
symmetric about the origin. The particle distribution would remain so
for all time (thus preventing the instability) if it were not for
truncation errors in the properties of the particles as determined
during the simulation.
In order to investigate the situation without the confusion caused by the
instability, we ran a number of simulations in which the distribution
of bodies was kept point symmetric about the origin at all times. To
this end we treat the bodies as pairs, and for each pair a and b,
we enforce that $\bmath{x}_{a}+\bmath{x}_{b}=0$ and
$\bmath{v}_{a}+\bmath{v}_{b}=0$ in the initial conditions and after each
time step.

This is effectively the equivalent of simulations using an expansion
in spherical harmonics which exclude all the terms with odd $(l,m)$.
Such experiments were carried out by \cite{Sellwood2003}.

\begin{figure}
  \centerline{\resizebox{\hsize}{!}{\includegraphics{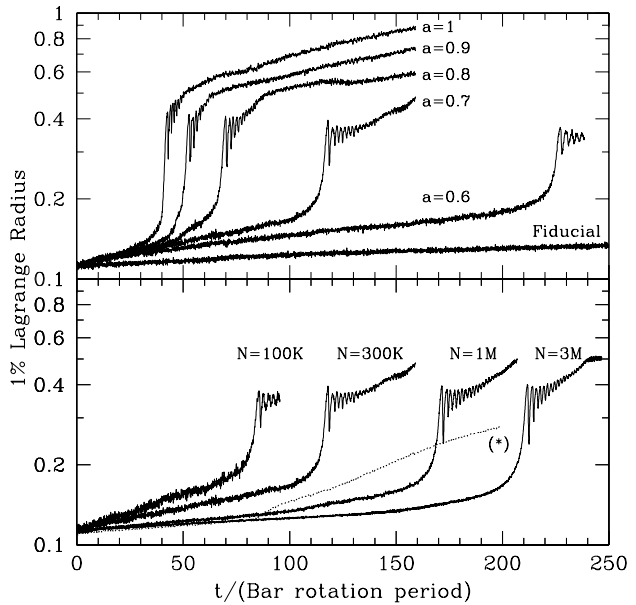}}}
  \caption{The 1\% Lagrange radius plotted against time for many different
    simulations.  \emph{Top}: the effect of increasing bar size with
    fixed $N=300\,000$ and enforced symmetry. Bar mass was defined as 
    30\% of the halo mass within the bar radius for all cases. Also shown are
    the results of a fiducial run with $N=300\,000$ and no imposed bar.
    \emph{Bottom}: the effect of increasing resolution (particle number), 
    for fixed bar size $a=0.7$ and enforced symmetry also showing \emph{(Dotted)} 
    simulation with conserved linear momentum with same bar size, 
    and $N=1\,000\,000$ (*) 
    \label{fig:varsize}}
\end{figure}

As Figure~\ref{fig:varsize} shows, the ability of a rotating bar to
remove a cusp is genuine, and not reliant upon instability. Even when
the simulation is completely symmetric, angular momentum transferred
from the bar to the halo eventually removes the cusp. The dependence
of the effect on $N$ and bar strength are comparable to the results
obtained by Sellwood (Figure~\ref{fig:varsize}). The major difference
to note is that our results do not yet seem to converge at high $N$.
This is likely caused by our force solver allowing stochastic jitter 
in the body distribution to be passed to the forces more readily than
a method involving only low-order spherical harmonics does.

\subsubsection{Using a more realistic bar model} \label{sec:results:stable:mom}

An alternative method for stabilising the simulation is to allow the bar 
to ``have mass'', and thus provide a restoring force on the halo. 
This was achieved in our simulations by 
including the monopole component of the bar potential in the force 
calculation. The initial distribution function of the halo with the 
additional monopole potential can be determined using Eddington's 
formula. In these simulations the monopole component
of the potential was included at all times, the quadrupole was added
adiabatically as in all previous simulations. Any off-centring of the halo 
causes an artificial transfer of linear momentum from the bar to the halo, 
so in addition the bar was constrained to move 
in such a way that the linear momentum of the bar-halo system was conserved.

Figure \ref{fig:varsize} shows a comparison of the results from this approach 
to those from the symmetrised simulations. The movement required of the bar 
in order to conserve the linear momentum of the system was extremely small, 
and only carried the bar a distance $\sim$$10^{-5} r_{s}$, which is several 
orders of magnitude smaller than the movement of the cusp seen in 
Figure~\ref{fig:quad}. Simulations which included the monopole in the bar 
potential, but in which the bar was pinned to the origin produced results 
which were nearly indistinguishable from those with a moving bar. It is clear 
that the restoring force provided by inclusion of the monopole term is 
sufficient to prevent the growth of the instability in this case. The 
presence of the monopole and removal of the artificially enforced 
symmetry prevent the rapid ``runaway'' evolution seen at $t\simeq170$ 
in the corresponding symmetrised simulation. However the bar still 
removes the cusp of the halo over approximately the same timescale 
in the two simulations.


\subsection{Angular Momentum conservation} \label{sec:results:angmom}
These experiments are deliberately kept as simple as possible, which
leaves them unrealistic. One major difference
between the simulations described thus far and the true situation is
the apparent infinite supply of angular momentum for the bar, which
keeps it rotating with the same pattern speed for the entirety of the
simulation, however much angular momentum has been transferred to the
halo. \cite{WeinbergKatz2002} pointed to a ``suite of simulations with
a slowing bar, whose pattern speed follows from the conservation of
angular momentum of the combined bar-halo system'' as further evidence
pointing to the effectiveness of this process in removing
the cusp. However these results are tainted by the instability of the
system, so we repeat the experiments with a slowing bar and the
symmetrisation method to prevent instability. This is somewhat similar
to experiments performed by \cite{Sellwood2003}, with the $l=1$
spherical harmonic suppressed.  Since initially the bar has zero mass,
and the mass (and thus angular momentum) is increased adiabatically,
angular momentum is introduced to the system by the growing bar. We
calculate the angular momentum of the halo after each time step and
use that to calculate the change in the pattern speed of the bar at
every point, so that the total angular momentum added to the halo is
no more than that of the bar at its full mass, rotating with its
original pattern speed.

\begin{figure}
  \centerline{\resizebox{\hsize}{!}{\includegraphics{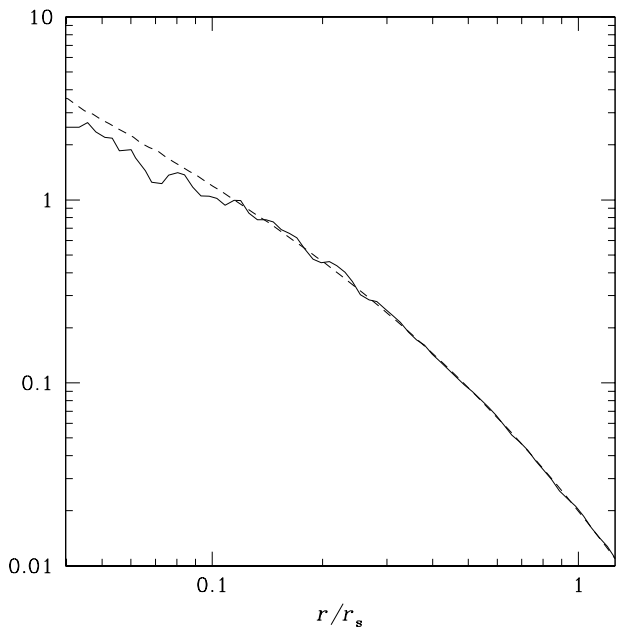}}}
  \caption{Density profile for a slowing bar in a $N=300\,000$ 
    representation of a Hernquist halo initially
    (\emph{dashed}) and after 16 initial bar rotation periods
    (\emph{solid}). The bar has not gone through 16 rotations, because
    it has decelerated, and by this point is rotating at 10\% of its
    initial pattern speed.\label{fig:slowbar}}
\end{figure}

As Figure~\ref{fig:slowbar} shows, when the bar is not given an
infinite supply of angular momentum, the situation is somewhat
different. Essentially all the angular momentum of the bar is absorbed
by the halo within 15 initial rotation periods (in fact the bar has
made ten complete revolutions at that point). This absorbed angular
momentum is insufficient to remove matter from the cusp to any great
extent. In experiments carried out without symmetrisation, the central
density did fall to approximately that seen in the experiments with a
bar rotating with a constant pattern speed. This was again because the
cusp of the distribution shifted from the origin.

\section{Conclusions} \label{sec:conclude}

We find that over-simplifying the interaction of a barred galaxy with
a cusped CDM halo may lead to substantial over-estimation of
the reduction in central density. The cause is an instability of the
simplified model: a slight off-centring of the CDM cusp from
the origin results in a net force from the bar, which is fixed to stay
centred, further driving the cusp away from the origin. The
instability is implicit in the model and is not simply a result of
determining the potential as a spherical harmonic expansion about the
origin. It seems very likely that the rapid evolution reported by
\cite{WeinbergKatz2002} and Sellwood (\citeyear[][when odd $l$ terms
where included in the computation of the forces]{Sellwood2003}) were
an artifact of this instability in conjunction with averaging over
spherical shells concentric with the origin (rather than the density
centre).

If this artificial instability is suppressed, either by enforcing reflection
symmetry w.r.t.\ the origin, or by including the monopole of the bar potential 
and conserving linear momentum, the removal of the CDM cusp is a
very slow process, requiring $\mathcal{O}(100)$ bar-rotations, i.e.\ a
significant fraction of a Hubble time. Moreover, this simplistic model
contains an additional artifact: the infinite supply of angular
momentum from the ever torquing bar. If this is removed by controlling
the total angular momentum, the cusp is at most slightly
modified. This means that, under these conditions, 
the amount of angular momentum needed to
remove a CDM cusp is far larger than that present in the inner
parts of disk galaxies.

Of course, this work makes a large number of assumptions. 
The halo has been assumed to be spherically symmetric, 
isotropic, without substructure, isolated and initially in equilibrium. 
The bar was assumed to be rigid, 
and gas physics and bar-disk interaction were neglected. These are the same 
assumptions made by the paper that proposed this mechanism 
\citep{WeinbergKatz2002}. These simulations cannot tell the whole story, 
but they do provide an insight into the fundamental physical mechanism.

We conclude that, contrary to the original proposal, these simulations 
suggest that angular momentum (and energy) transport from
a disk-bar is not an effective way to destroy CDM cusps, in
agreement with studies using less simplistic models
\citep{ValenzuelaKlypin2003, Athanassoula2004}. If galactic scale CDM halos 
truly are cored, it seems that some other process is responsible.

\section*{acknowledgement}
The authors would like to thank their referee, Kelly Holley-Bockelmann, for her
constructive suggestions.
PJM acknowledges the support of the UK Particle Physics and Astronomy 
Research Council (PPARC) through a research student fellowship. 
Astrophysics research at the University of Leicester is also supported 
through a PPARC rolling grant.


\begin{thebibliography}{}

\bibitem[\protect\citeauthoryear{{Athanassoula}}{{Athanassoula}}{2004}]{Athana%
ssoula2004}
{Athanassoula} E.,  2004, in IAU Symposium Vol.~220, {Bars and the connection
  between dark and visible matter}.
p.~255

\bibitem[\protect\citeauthoryear{{Barnes} \& {Hut}}{{Barnes} \&
  {Hut}}{1986}]{BarnesHut1986}
{Barnes} J.,  {Hut} P.,  1986, Nature, 324, 446

\bibitem[\protect\citeauthoryear{{Binney}, {Gerhard} \& {Silk}}{{Binney}
  et~al.}{2001}]{BinneyGerhardSilk2001}
{Binney} J.,  {Gerhard} O.,    {Silk} J.,  2001, MNRAS, 321, 471

\bibitem[\protect\citeauthoryear{{Binney} \& {Evans}}{{Binney} \&
  {Evans}}{2001}]{BinneyEvans2001}
{Binney} J.~J.,  {Evans} N.~W.,  2001, MNRAS, 327, L27

\bibitem[\protect\citeauthoryear{{Blais-Ouellette}, {Amram} \&
  {Carignan}}{{Blais-Ouellette} et~al.}{2001}]{BlaisOuelletteAmramCarignan2001}
{Blais-Ouellette} S.,  {Amram} P.,    {Carignan} C.,  2001, AJ, 121, 1952

\bibitem[\protect\citeauthoryear{{Blumenthal}, {Faber}, {Flores} \&
  {Primack}}{{Blumenthal} et~al.}{1986}]{BlumenthalEtal1986}
{Blumenthal} G.~R.,  {Faber} S.~M.,  {Flores} R.,    {Primack} J.~R.,  1986,
  ApJ, 301, 27

\bibitem[\protect\citeauthoryear{{Bode}, {Ostriker} \& {Turok}}{{Bode}
  et~al.}{2001}]{BodeOstrikerTurok2001}
{Bode} P.,  {Ostriker} J.~P.,    {Turok} N.,  2001, ApJ, 556, 93

\bibitem[\protect\citeauthoryear{{C{\^ o}t{\' e}}, {Carignan} \&
  {Freeman}}{{C{\^ o}t{\' e}} et~al.}{2000}]{CoteCarignanFreeman2000}
{C{\^ o}t{\' e}} S.,  {Carignan} C.,    {Freeman} K.~C.,  2000, AJ, 120, 3027

\bibitem[\protect\citeauthoryear{{Col{\'{\i}}n}, {Avila-Reese} \&
  {Valenzuela}}{{Col{\'{\i}}n} et~al.}{2000}]{ColinAvilaReeseValenzuela2000}
{Col{\'{\i}}n} P.,  {Avila-Reese} V.,    {Valenzuela} O.,  2000, ApJ, 542, 622

\bibitem[{{de Blok} {et~al.}(2001{\natexlab{a}}){de Blok}, {McGaugh}, \&
  {Rubin}}]{deBlokMcGaughRubin2001}
{de Blok}, W.~J.~G., {McGaugh}, S.~S., \& {Rubin}, V.~C. 2001{\natexlab{a}},
  AJ, 122, 2396
 
\bibitem[{{de Blok} {et~al.}(2001{\natexlab{b}}){de Blok}, {McGaugh}, {Bosma},
  \& {Rubin}}]{deBlokEtal2001}
{de Blok}, W.~J.~G., {McGaugh}, S.~S., {Bosma}, A., \& {Rubin}, V.~C.
  2001{\natexlab{b}}, ApJl, 552, L23
 
\bibitem[\protect\citeauthoryear{{Dehnen}}{{Dehnen}}{1993}]{Dehnen1993}
{Dehnen} W.,  1993, MNRAS, 265, 250

\bibitem[\protect\citeauthoryear{{Dehnen}}{{Dehnen}}{2000}]{Dehnen2000:falcON}
{Dehnen} W.,  2000, ApJl, 536, L39

\bibitem[\protect\citeauthoryear{{Dehnen}}{{Dehnen}}{2002}]{Dehnen2002}
{Dehnen} W.,  2002, J.~Comp.~Phys., 179, 27

\bibitem[\protect\citeauthoryear{{Eddington}}{{Eddington}}{1916}]{Eddington1916}
{Eddington} A.~S., 1916, MNRAS, 76, 572

\bibitem[\protect\citeauthoryear{{El-Zant}, {Shlosman} \& {Hoffman}}{{El-Zant}
  et~al.}{2001}]{ElZantShlosmanHoffman2001}
{El-Zant} A.,  {Shlosman} I.,    {Hoffman} Y.,  2001, ApJ, 560, 636

\bibitem[\protect\citeauthoryear{{Ferrers}}{{Ferrers}}{1877}]{Ferrers1877}
{Ferrers} N.~M.,  1877, Q.~J.~Pure Appl.~Math., 14, 1

\bibitem[\protect\citeauthoryear{{Gentile}, {Salucci}, {Klein}, {Vergani} \&
  {Kalberla}}{{Gentile} et~al.}{2004}]{GentileEtal2004}
{Gentile} G.,  {Salucci} P.,  {Klein} U.,  {Vergani} D.,    {Kalberla} P.,
  2004, MNRAS, 351, 903

\bibitem[\protect\citeauthoryear{{Hayashi}, {Navarro}, {Power}, {Jenkins},
  {Frenk}, {White}, {Springel}, {Stadel} \& {Quinn}}{{Hayashi}
  et~al.}{2004}]{HayashiEtal2004}
{Hayashi} E.,  {Navarro} J.~F.,  {Power} C.,  {Jenkins} A.~R.,  {Frenk} C.~S.,
  {White} S.~D.~M.,  {Springel} V.,  {Stadel} J.,    {Quinn} T.~R.,  2004,
  MNRAS, 355, 794

\bibitem[\protect\citeauthoryear{{Hernquist}}{{Hernquist}}{1990}]{Hernquist199%
0}
{Hernquist} L.,  1990, ApJ, 356, 359

\bibitem[\protect\citeauthoryear{{Hernquist} \& {Ostriker}}{{Hernquist} \&
  {Ostriker}}{1992}]{HernquistOstriker1992}
{Hernquist} L.,  {Ostriker} J.~P.,  1992, ApJ, 386, 375

\bibitem[\protect\citeauthoryear{{Hernquist} \& {Weinberg}}{{Hernquist} \&
  {Weinberg}}{1992}]{HernquistWeinberg1992}
{Hernquist} L.,  {Weinberg} M.~D.,  1992, ApJ, 400, 80

\bibitem[\protect\citeauthoryear{{Holley-Bockelmann}, {Weinberg} \& {Katz}}{{Holley-Bockelmann} et~al.}{2003}]{HolleyBockelmannWeinbergKatz2003}
  {Holley-Bockelmann} K., {Weinberg} M.~D., {Katz} N., 2003 astro-ph/0306374

\bibitem[\protect\citeauthoryear{{Kaplinghat}, {Knox} \& {Turner}}{{Kaplinghat}
  et~al.}{2000}]{KaplinghatKnoxTurner2000}
{Kaplinghat} M.,  {Knox} L.,    {Turner} M.~S.,  2000, Physical Review Letters,
  85, 3335

\bibitem[\protect\citeauthoryear{{Moore}, {Governato}, {Quinn}, {Stadel} \&
  {Lake}}{{Moore} et~al.}{1998}]{MooreEtal1998}
{Moore} B.,  {Governato} F.,  {Quinn} T.,  {Stadel} J.,    {Lake} G.,  1998,
  ApJl, 499, L5+

\bibitem[\protect\citeauthoryear{{Navarro}, {Frenk} \& {White}}{{Navarro}
  et~al.}{1996}]{NavarroFrenkWhite1996}
{Navarro} J.~F.,  {Frenk} C.~S.,    {White} S.~D.~M.,  1996, ApJ, 462, 563

\bibitem[\protect\citeauthoryear{{Power}, {Navarro}, {Jenkins}, {Frenk},
  {White}, {Springel}, {Stadel} \& {Quinn}}{{Power}
  et~al.}{2003}]{PowerEtal2003}
{Power} C.,  {Navarro} J.~F.,  {Jenkins} A.,  {Frenk} C.~S.,  {White} S.~D.~M.,
   {Springel} V.,  {Stadel} J.,    {Quinn} T.,  2003, MNRAS, 338, 14

\bibitem[\protect\citeauthoryear{{Salucci}, {Walter} \& {Borriello}}{{Salucci}
  et~al.}{2003}]{SalucciWalterBorriello2003}
{Salucci} P.,  {Walter} F.,    {Borriello} A.,  2003, Aap, 409, 53

\bibitem[\protect\citeauthoryear{{Sellwood}}{{Sellwood}}{2003}]{Sellwood2003}
{Sellwood} J.~A.,  2003, ApJ, 587, 638

\bibitem[\protect\citeauthoryear{{Sellwood}}{{Sellwood}}{2005}]{Sellwood2005}
{Sellwood} J.~A.,  2005, ApJ, submitted (astro-ph/0407533)

\bibitem[\protect\citeauthoryear{{Spergel} \& {Steinhardt}}{{Spergel} \&
  {Steinhardt}}{2000}]{SpergelSteinhardt2000}
{Spergel} D.~N.,  {Steinhardt} P.~J.,  2000, Physical Review Letters, 84, 3760

\bibitem[\protect\citeauthoryear{{Swaters},{Madore},{van den Bosch} \& {Balcells}}{{Swaters} et~al.}{2003}]{SwatersEtal2003}
{Swaters} R.~A., {Madore} B.~F., {van den Bosch} F.~C., {Balcells} M., 2003, ApJ, 583, 732

\bibitem[\protect\citeauthoryear{{Trott} \& {Webster}}{{Trott} \&
  {Webster}}{2002}]{TrottWebster2002}
{Trott} C.~M.,  {Webster} R.~L.,  2002, MNRAS, 334, 621

\bibitem[\protect\citeauthoryear{{Valenzuela} \& {Klypin}}{{Valenzuela} \&
  {Klypin}}{2003}]{ValenzuelaKlypin2003}
{Valenzuela} O.,  {Klypin} A.,  2003, MNRAS, 345, 406

\bibitem[\protect\citeauthoryear{{van den Bosch} \& {Swaters}}{{van den Bosch}
  \& {Swaters}}{2001}]{vandenBoschSwaters2001}
{van den Bosch} F.~C.,  {Swaters} R.~A.,  2001, MNRAS, 325, 1017

\bibitem[\protect\citeauthoryear{{Weinberg}}{{Weinberg}}{2004}]{Weinberg2004}
{Weinberg} M.~D.,  2004, astro-ph/0404169

\bibitem[\protect\citeauthoryear{{Weinberg}}{{Weinberg}}{1999}]{Weinberg1999}
{Weinberg} M.~D.,  1999, AJ, 117, 629

\bibitem[\protect\citeauthoryear{{Weinberg} \& {Katz}}{{Weinberg} \&
  {Katz}}{2002}]{WeinbergKatz2002}
{Weinberg} M.~D.,  {Katz} N.,  2002, ApJ, 580, 627

\bibitem[\protect\citeauthoryear{{White}}{{White}}{1983}]{White1983}
{White} S.~D.~M.,  1983, ApJ, 274, 53

\end{thebibliography}

\end{document}